\documentclass[12pt,preprint]{aastex}
\usepackage{emulateapj5}

\shorttitle{Cold Dark Matter's Small Scale Crisis Grows Up}
\shortauthors{D'Onghia $\&$ Lake}

\begin{document}

\title{Cold Dark Matter's Small Scale Crisis Grows Up }

\author{Elena D'Onghia\altaffilmark{1,2}\\ }
\affil{Max-Planck-Institute for Astronomy, 
D-69117 Heidelberg, Germany}
\email{donghia@usm.uni-muenchen.de}

\author{George Lake\altaffilmark{3}\\}
\affil{Department. of Physics, Washington State University, 
Pullman WA  99164-2814}
\email{lake@wsu.edu}

\altaffiltext{1}{Max-Planck-Institut f\"ur extraterrestrische Physik,    
85748 Garching, Germany}
\altaffiltext{2}{Institut f\"ur Astronomie und Astrophysik,   
Scheinerstrasse 1 81679 Munich, Germany}
\altaffiltext{3}{Chief Scientist, Arctic Region 
Supercomputer Center Box 756020 Fairbanks, AK 99775}

\begin{abstract}
The Cold Dark Matter (CDM) theory predicts a wealth 
of substructure within dark halos. 
These predictions match observations of galaxy clusters 
like the nearby Virgo cluster. However, 
CDM has a ``small scale crisis" since galaxies 
dominate  
the halo with little substructure while the model 
predicts that galaxies  should be scaled versions of
galaxy clusters with abundant substructure. 
Compared to CDM predictions,  the Milky Way and 
Andromeda are ``missing"  objects with velocity 
dispersions $\sigma \geq 10$ km~s$^{-1}$.
The energy scale of these missing satellites is 
low enough that stellar winds and supernovae 
might remove gas and
suppress the formation of their luminous stellar components.  
Here, we show
that the small scale crisis persists in fossil groups 
that have masses
of up to 40\% of the nearby Virgo cluster of galaxies.
Fossil groups are missing  satellites with
luminosities that occur at the predicted frequency in the
Virgo cluster. Moreover, the ``missing galaxies" in fossil 
groups are nearly as luminous as the Milky Way with a velocity 
dispersion
$\sigma \leq$ 150  km~s$^{-1}$.

\end{abstract}

\keywords{cosmology: observations -- cosmology: -- 
dark matter -- galaxies: clusters: general -- galaxies: 
formation}

\section{Introduction}
The formation of structure in the universe by the hierarchical 
clustering is an elegant 
and well-defined theory that 
explains observations 
of the universe on large scales (Blumenthal et al. 1984).
In early simulations, it seemed that 
merging was too efficient
to be consistent with the observed hierarchy
of structures (White $\&$ Rees 1978). This {\it ``overmerging''} problem 
was reproduced in simulations for several years  
(White et al. 1987, Frenk et al. 1988).   
While overmerging was a virtue on the 
scale of galaxies, it was a problem for rich clusters 
of galaxies.  Solutions focused on the role of gas 
dynamics in making lumps within rich clusters of 
galaxies (Katz $\&$ White 1993).
Eventually, Moore, Katz and Lake (1996) showed that numerical 
heating dominated over physical mechanisms
unless simulations had nearly $10^6$ particles within 
the virial radius of a cluster.  Simulations with
this resolution reversed the picture,
overmerging disappeared and halos the size 
of the Milky Way are predicted to have 
nearly the same scaled distribution of substructure as 
the Virgo cluster (Moore et al. 1999, 
hereafter M99; Klypin et al 1999). 

This strong prediction can be tested
observationally. A Milky Way sized halo should have 
$\sim$500 satellites within 500 kpc, with circular 
velocities greater than 5\% of the parent halo's  
velocity, {\it i.e.} $V_{cir}/V_{parent}>0.05$, in contrast 
to a scant 11 that are
observed (Klypin et al. 1999; M99).

It has been suggested that the stellar components of
the  Milky Way satellites might have accumulated in the 
core regions of their dark halos where the characteristic 
velocities $\sigma$ are smaller than the
asymptotic value of $V_{cir}$.  The observed velocities of 
Milky Way's satellites would be re-mapped to much higher 
peak values than expected, shifting the objects plotted in 
Fig.1 (left panel) to the right until they match the 
theoretical prediction (Hayashi et al. 2003). There are still 
many satellites missing at lower peak velocities compared to 
CDM predictions, but these are declared to have gone dark 
owing to the ejection of gas from systems with low escape 
velocities of only 20 - 60 km~s$^{-1}$.  These 
objects are also deficient in the field 
({\it c.f.} Kauffmann, White $\&$ Guiderdoni 1993) where the same 
processes could keep them from being observed.

Is this the solution to the overmerging crisis? The ROSAT X-ray 
satellite discovered a new class of objects: 
fossil groups (Ponman et al. 1994).
RXJ1340.6+4018 at redshift $0.171$ is the 
archetype with a bright isolated elliptical 
galaxy $M_{R}=-22.7$, surrounded by dark matter
and a hot gaseous halo.  The spatial extent of the X-ray 
emission, $\sim 500$ kpc, the total mass, $\sim 6 \cdot 10^{13}$ 
M$_{\odot}$,
and the mass of the hot gas correspond to a galaxy cluster 
$\sim 40$\% as massive as Virgo, and the optical
luminosity of the central galaxy is comparable to that 
of cluster cD galaxies (Jones, Ponman $\&$ Forbes 2000, 
hereafter JPF00).
Five additional fossil groups have been confirmed 
spectroscopically.  For one of them, RXJ1416.4 
the X-rays temperature is estimated to be $\sim 1.5$ keV 
(Jones et al. 2003).
Fossil groups are not rare. Their number density is 
$\sim 2.4 \cdot 10^{-7}$ $h_{50}^{3}$ Mpc$^{-3}$ using the 
definition that they have a dominating giant elliptical galaxy 
with the next brightest object being 2 magnitudes fainter, 
embedded in a X-ray halo with a luminosity
10-60 \% of the Virgo cluster (Vikhlinin et al. 1999; 
Jones et al. 2003).
They comprise $\sim$20\% of all clusters and groups with 
an X-ray luminosity larger than $2.5 \cdot 10^{42} h_{50}^{-2}$ 
ergs$^{-1}$, and host nearly all field galaxies 
brighter than $M_{R}=-22.5$ (Vikhlinin et al 1999).
Their total mass density is comparable to massive galaxy 
clusters. Their high mass-to-light ratio, $M/L_R \sim 300$, is
comparable to Virgo.  The luminosity-temperature relations are 
also similar (Jones et al. 2003).

We define {\it overmerged} systems as objects dominated by a 
single central object with weak substructure with the Milky Way 
as a local prototype. In contrast, {\it clusters of galaxies} 
have abundant substructure and  
a central galaxy with a velocity dispersion that is considerably 
less than the overall dark halo, the local prototype being the 
Virgo cluster.
In this Letter, we examine overmerging in systems 
with masses intermediate between the Milky Way and the Virgo 
cluster.  

\section{The cumulative substructure function in Fossil Groups}

We compare the cosmological model predictions (De Lucia et al. 2004)
to the substructure function of RXJ1340.6+4018, Virgo and Coma clusters of galaxies,
Hickson Compact Groups (HCGs) and the Local Group. In Fig.1 
the cumulative substructure function is the number of objects 
with velocities greater than a fraction of the parent halo's 
velocity.

For the groups and clusters, we convert 
luminosity functions (LFs) to substructure functions using 
using the Tully-Fisher relation for the spirals (Tully \& Pierce 2000)
and the Faber-Jackson (1976) relation for 
early-type galaxies.
RXJ1340.6+4018 has a velocity dispersion of 
$\sigma_{parent} \sim 380$ km ~s$^{-1}$  
($V_{parent}=\sqrt{2} \sigma_{parent}$) and its brightest 
galaxy has $\sigma \sim 260$ km~ s$^{-1}$ (JPF00). The 
substructure plots skip the largest central galaxy. With the 
observations, it can be difficult to disentangle the central 
object from diffuse light in the cluster.  However, a greater 
uncertainty comes from the simulations that might still have 
too little resolution in the very center of the cluster (Taylor, Silk and Babul 2003).  
For the  second brightest object,  
we find $V_{cir}/V_{parent} \sim 0.35$ and proceed down the
LF to construct a substructure function.
For the Virgo cluster,
we use the LF of 
Binggeli, Sandage $\&$ Tammann (1985) and for the Coma cluster the LF of 
Trentham (1998).

The cumulative distribution of satellites in the Milky Way's 
halo and Andromeda's halo are also plotted.  Here, 
the measured one-dimensional velocity dispersions of satellites
(Mateo 1998)  are converted to circular velocities assuming an 
isotropic velocity dispersion (M99).

We would like to have a sample of LFs for 
objects with intermediate mass between the Local Group 
and Virgo cluster.   There are only a few LFs
known in this range, most of them from studies of Hickson 
(1982) compact groups.
Hunsberger, Charlton $\&$ Zaritsky (1998)
constructed an LF from 39 compact groups.  To convert this to the
substructure function in Fig.1 (right panel),  
we adopt $\sigma \sim 370$ km~s$^{-1}$ for the typical 
velocity of the parent's halo and use the Tully-Fisher relation that
Mendes de Oliveira et al. (2003) have shown applies to galaxies in HCGs.

In the right panel of Fig.1, we show a composite substructure function for the  5 loose groups
in  Zabludoff $\&$ Mulchaey (2000). 
These look very different than the other substructure functions, 
appearing to be shifted strongly to the right compared to the composite for the 39 HCGs.
While these objects could be very 
different,  loose groups are likely to have even more contamination than typical
HCGs (Hernquist, Katz and
Weinberg 1995).    Zabludoff $\&$ Mulchaey (2000) point to the Local Group as an archetype 
of loose groups.   While the Local Group is certainly a physical association, it is not bound and virialized.  
If we treated it as a group, the Milky Way would appear as the second brightest member
and the rest of the points would move upward by a factor of 2.  This would indeed be an archetypal
substructure function for a loose group.   If instead,  we wait for the virialization of the group and 
the merger of M31 and the Milky Way, we would see something extremely similar to
the substructure function of the individual virialized systems. 
While the fist few points of the combined substructure function of the
39 HCGs
place them high on the substructure function, the LFs
quiclky flatten at the faint end and show a deficit
of structure there (right panel Fig.1).  The combined substructure
function
of the 5
loose groups from  Zabludoff $\&$ Mulchaey (2000) show behavior that is
intermediate
between the HCGs and the brighter clusters.

Zabludoff  $\&$ Mulchaey (1998b) find that their two groups with the greatest number
of members (HCG 62 and NGC 741) can be broken into two distinct subgroups.  They suspect 
that the fraction with such structure is much higher than 40\% since their statistic is less sensitive
for the groups with fewer members and requires an angular offset of the centroids of the subclumps.

In the left panel of Fig.1, the similarity of the substructure 
function in RXJ1340.6+4018 
to the Milky Way and Andromeda is striking. It shows that fossil 
groups are also {\it overmerged} objects.
However, for galaxies of any given $V_{cir}$ that are missing in 
the fossil group, galaxies with the same $V_{cir}$ appear with the 
predicted frequency in Virgo and are observed in the field as well. 
The Virgo cluster contains six L* galaxies (Binggeli, Sandage 
$\&$ Tammann 1985) with 
L* being a characteristic luminosity in the luminosity function 
and is roughly the luminosity of the Milky Way. 
Fossil groups show one or no L* galaxies (Mulchaey $\&$ Zabludoff 
1999; Jones et al. 2003), while the CDM substructure function would predict a few in each group.
The likelihood that the substructure in fossil groups and in 
the Virgo cluster is drawn from the same, universal cosmological 
distribution function is negligibly small, especially at the 
low mass end.

\section{The transition from Overmerging to Galaxy Clusters} 
Where does the transition from {\it overmerged} systems to 
galaxy clusters with substantial substructure occur?  
Is the transition from overmerging
to clusters  smooth, abrupt or merely ill determined with a 
scattering of points?  

Studies of Hickson (1982) Compact Groups, loose groups (Zabludoff  $\&$ Mulchaey 1998a,b), 
the 2dF Galaxy Redshift Survey (2dFGRS; Colless et al. 2001),   
the Sloan Digital Sky Survey (SDSS; York et al. 2000) are sources for catalogs
of groups and clusters.  
Balogh et al. (2003) analyzed both surveys to look at ``galaxy ecology" or star formation as a function of
environment.  Desai et al. (2003)  fit some circular velocity functions to a sample from the 
SDSS and
compared this to the large simulation by Reed et al. (2003).  The critical range of group velocity 
dispersions is  $250-400$ km s$^{-1}$, as this is  where one would like to see the transition from 
overmerging on the scale of galaxies to the abundant substructure in clusters.  In this range, 
there are 39 objects in the HCG sample of 
(Hunsberger, Charlton $\&$ Zaritsky 1998), 5 loose groups in the sample of Zabludoff  $\&$ Mulchaey (1998a), 
9 SDSS groups from Desai et al. (2003) and roughly 40 groups from the 2dFGRS 
(Balogh et al. 2003).   Only the HCGs and the loose groups have LFs that are deep enough to be
used in the bottom panel of Fig.2.  There are a few other sources with not quite enough information to be used.
For example, the LFs observed by Muriel, Valotto $\&$ Lambas (1998) shows a number of groups with 
LFs that are similar to the HCG sample of Hunsberger, Charlton $\&$ Zaritsky 1998, but the typical 
velocity dispersions for the groups is unknown.
 
We plot 
the number of galaxies
brighter than $M_B < -19$ versus velocity dispersion in the top panel of Fig. 2 using the data from
Balogh et al. (2003) and  Desai et al. (2003).  As expected,  the number
of galaxies increases with the velocity dispersion of the group.  We added a line that
shows what one would observe if the substructure function of Virgo were universal.  
At low dispersion,
it appears that objects have more substructure than scaling Virgo.  This owes to the criteria that there must be 10
members to be included as a group in the samples.  With the cut of  $M_B < -19$, the
Milky Way would still be consistent with a scaled Virgo substructure function.  At the high velocity 
dispersion end, 
there is less substructure than the scaled Virgo substructure function predicts.  This is consistent with 
previous studies where
the luminosity function within large clusters was relatively constant rather than scaling
with cluster size (De Propis et al. 2003).  There is not a large sample of such clusters in CDM simulations.  
There are a few high resolution runs of individual clusters (Borgani et al. 2002) and the high resolution run of 
Reed et al. (2003) simulated a volume of 100 Mpc side which is not large enough for a good sample 
of large clusters.    

There are fewer mass functions that reach $V_{cir}/V_{parent}>0.05$ and these have been 
collected in the bottom panel
of Fig. 2. 
This sample includes all the systems shown in Fig. 1 and adds 
the Fornax cluster.  Fornax has a velocity dispersion  $\sigma \sim 374$ km~s$^{-1}$, comparable to the RXJ1340.6, 
but has little diffuse X-ray emission (e.g. Horner, Mushotzky $\&$ Scharf 1999) and considerably
more substructure (obtained from the luminosity function of Ferguson $\&$ Sandage 
1989), albeit less than a scaled version of Virgo would predict.
For RXJ1340.6+4018, integrating the luminosity function 
within the large error bars gives an 
upper limit of $\sim 30$ members with $V_{cir}/V_{parent}>0.05$, 
but only nine are spectroscopically confirmed.  
We use 9 as the number of substructures and show the current 
uncertainty with an error bar to 30 in Fig.2.  

The loose groups might not be a single bound systems but 
projections of filaments of galaxies (Hernquist, Katz $\&$ 
Weinberg 1995) or superposition of multiple structures (Zabludoff $\&$ Mulchaey 1998b). 
At the moment, the loose groups are the main objects that we have 
have in the transition region intermediate between overmerged  systems and clusters.
Fig.2 is sparsely populated, but argues for substantial variation of properties of
systems with velocity dispersions of $300-400$ km s$^{-1}$.  

\section{Discussion}
What is the origin of the fossil groups?
The similarity in their cumulative galaxy distribution with
the Local Group (Fig. 1, left panel) suggests that they are 
the end result of merging of L* galaxies in low density 
environments (Jones et al. 2003).
The giant elliptical in 
RXJ1340.6+4018 has no spectral features which would indicate 
recent star formation. Hence, the last major merger must have 
occurred several gigayears ago (JPF00).  

Although the substructure function of
fossil groups and the  Local Group are similar, the merger of 
the Milky Way and the Andromeda galaxy will not form an 
X-ray dominated fossil group. The mass of the merged Local Group
will be  $\approx 3-5\cdot 10^{12}$M$_{\odot}$ 
(Kahn $\&$ Woltjer 1959) within 300 kpc with 
$V_{max}\approx 290$ km~s$^{-1}$, 10\%
higher  than Andromeda and significantly less than observed
fossil groups. Merging won't change the circular velocity of the
satellites,  though they will change morphology and than will fade. 
The result will look more like Centaurus A which has a 
substructure function 
like the Local Group with an elliptical at the middle but no 
X-ray emission  and a total mass that is less than fossil groups
(Karachentsev et al. 2002).
Additional  ``two by two" hierarchical mergering would make a system that 
matched the optical properties of a fossil group,  but
it is not clear why ``overmerging" would propagate up the 
hierarchy to produce fossil groups while clusters like Virgo have 
galaxies of the same luminosity as those that are
missing in fossil groups.  

X-ray halos in fossil groups and clusters are an outstanding problem 
(Mulchaey 2000).  In general, it is difficult to keep all the gas from cooling at
early times and becoming a part of the galaxies.
Since Virgo has extensive 
X-ray emission while Fornax has a 
paltry intracluster medium, we have all 4 combinations of 
systems that are  overmerged or having abundant substructure 
together with those with
abundant  or very little intracluster medium.
It might  well be that having fossil groups among the progenitors 
of a cluster is key to producing their X-ray emission.  

Dynamical friction and merging aren't 
a general solution to the overmerging  problem.
Clearly, 
these dynamical effects were included in the full numerical 
simulations that first highlighted  the problem in the CDM model.
Any specific substructure function evolves in the 
same way by dynamical friction  and merging independent of the 
parent mass.  The dynamical friction timescale 
$t_{df}$ is proportional to the crossing time of a system 
$t_{cr}$ divided by the fractional 
mass of the sinking object (e.g. 
$t_{df} \sim 0.05 \  t_{cr}/(M_{sinker}/M_{parent})$.   
The crossing time of all virialized 
halos is the same.  Further the fractional mass of the 
sinking satellite is a function of  the variable 
$V_{cir}/V_{parent}$ in the substructure function and the 
tidal radius of the satellite which is determined by it's 
orbital  pericenter as fraction of the virial 
radius $r_{peri}/r_{virial}$.  All of these quantities scale 
with parent mass such that the  evolution of the substructure 
function is independent of the mass of the parent halo.  
Dynamical friction can 
be important in promoting the merger of the largest objects 
in less than one Hubble time, within a 
parent halo, but dynamical friction alone will not create 
substructure functions
that are different for different parent masses.
Of course, galaxies could have had a long time to evolve 
by dynamical friction, but this will not affect the
substructure function below $V_{circ}/V_{parent}$ of 0.2.

What mechanisms could  explain the substructure function of
fossil groups? The first thought might be merely cosmic variance.
The top panel of Fig. 2 shows the variance in the groups selected from 
the SDSS (Desai et al. 2003).  
The figures in
Desai et al. (2003) show that the variance seen in the groups selected
from the SDSS (shown here
in Fig. 2) is already 2-3 times greater
than observed in the simulation of a (100Mpc)$^3$ volume simulated by
Reed et al. (2003).
The fossil groups lie well outside of the variation seen in the large simulation, but they
are rare enough that a larger volume is required to be definitive.

One might extend the proposal of Hayashi et al. (2003) who argue 
for  shifting points to the right and then blowing the baryons out of the smallest objects.
While gas ejection is attractive to explain the missing satellites of the 
Local Group, it would take nearly 10  times as much energy to 
blow the gas out of the missing galaxies in fossil groups.  
Further,  the gas ejection must also
be tuned to the environment since the same galaxies appear at 
the predicted  frequency in clusters and the field.  We have no 
evidence that L* galaxies are fragile in either of these 
environments.  The same  tuning argument is a severe constraint 
on solutions that alter the initial cosmic fluctuation spectrum.
Star formation could be 
suppressed at a higher  energy scale in fossil groups by 
appealing to intense bursts of star formation
as seen in starburst galaxies at high  redshift or by appealing 
to the power of a super-massive black-hole.
The comoving number densities of the ULIRG starburst galaxies at 
high redshift roughly matches the number densities of the halos 
predicted at the same redshift 
by hierarchical merging leading to the speculation that there is 
a starburst galaxy in every halo with mass 
$\sim 10^{13} M_{\odot}$
 (Somerville, Primack $\&$ Faber 2001).
 These halos will also be the progenitors
 of objects the size of fossil groups and larger.

Such an energy injection could also create the 
reservoir of gas needed for large clusters as well as the observed
entropy floor in that gas.  As substructure is suppressed, gas 
falling into the deep  potential well of a massive dark cluster
will be more effectively heated in an accretion shock.  This 
could enhance  substructure suppression and boost the fraction 
of baryons that settle into a single luminous galaxy. However,  
the presence of intracluster gas doesn't appear to be related 
to the  substructure function as we see all combinations of X-ray 
emission and substructure functions in our small sample of systems. 
It's also  not clear why energy input was so effective at 
suppressing  substructure in a  fossil group, but so ineffective 
in Fornax and Virgo which have the powerful sources Fornax A and 
M87.

Fossil  groups offer a better environment to test these hypotheses
than galaxies that are counterparts to the Milky Way. If the  
galaxies have been altered in fossil groups, they need to be a 
factor of ten fainter than expected from their  dark matter mass. 
As a result, they should have velocity
dispersions that are anomalously high by nearly a factor of two. 

Gravitational  lensing provides three possible tests. 
M99 suggested direct detection of mass clumps
using the  brightness ratio of lensed images.  
With this technique, Dalal and 
Kochanek  (2002) used a sample of isolated ellipticals that are 
likely the centers of  fossil groups.  They find that a few 
percent of the mass in halos is in clumps with masses in the 
range of $10^6 - 10^9$M$_{\odot}$  which is a factor of a few 
below what is seen in simulations (M99; Klypin et al. 1999; 
Ghigna et al. 2000).  However, 
Zentner and  Bullock (2003) found that Dalal and Kochanek's 
model underestimated the 
substructure  since they placed substructures uniformly in the 
halo rather than allowing for stripping and  destruction in the 
central regions where they had the greatest sensitivity.
There are other consequences associated with lumps that are 
more massive than $10^9$ M$_{\odot}$. In the strong lensing case, 
the positions of images will shift betraying
individual lumps rather than just their statistical properties.
This was not seen by Dalal and Kochanek (2002).
Shifts in the center of mass may also be seen with weak lensing 
maps. 
Here, one would compare the centers defined by the brightest
galaxy, the X-ray emission and the lensing map. If there is significant
clumping, the brightest galaxy will  be displaced from the center
of mass defined by the other two.  The center of
the X-ray emission generally agrees with the location of the
  brightest galaxy (Mulchaey 2000), but the lensing map should
be a more sensitive test.

 \section{Conclusions}
 
Gas processes have been invoked to explain
the absence of the dwarf satellites in the Milky Way and
the field compared to CDM 
predictions. However, we now find that overmerging persists to 
the larger mass scale of fossil groups where galaxies as massive 
as the Milky Way and the Large Magellanic Cloud are ``missing",
though they appear at the predicted abundance in both the field 
and clusters of galaxies. 
Fig.2 shows that the ``overmerging'' behavior is
more dependant on the mass of the parent halo rather than the mass
of the satellite, though this could owe to our limited sample size.
We have pointed to some key observations that can resolve whether 
this owes to energetic phenomena or is a result of an unknown 
source that is closely tied to the mass scale of the parent halo.
To resolve this issue, we need more systems with velocity dispersions
in the range of 300-400 km s$^{-1}$.  
For the present data, {\it overmerging} behaviour seems to be the 
generic behaviour for objects with T$\le 1$ keV .






 \acknowledgments

We would like to thank Simon White, Andi Burkert, Chris Kochanek and Claudia Mendes de Oliveira for useful
 discussions and comments.
G.L acknowledges support  from the US National Science Foundation.





\clearpage

\begin{figure}
\plotone{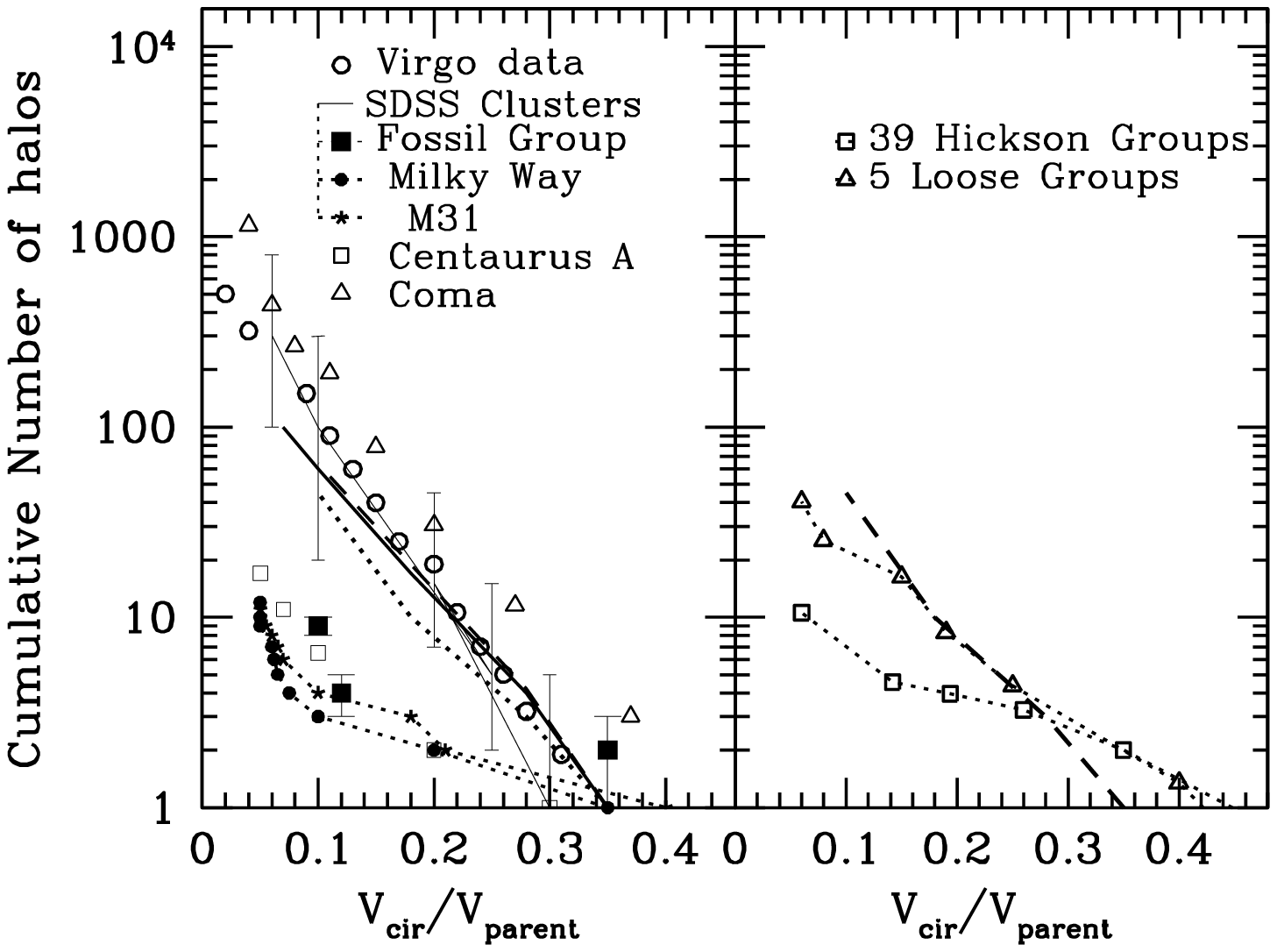}
\caption{The observed cumulative substructure function of 
galaxies within RXJ1340.6+4018 (fossil group), 
Virgo and Coma clusters of galaxies, clusters from SDSS, the Local Group and Centaurus A 
compared to CDM predictions (De Lucia et al. 2004) (left panel). 
The thick solid line is the CDM prediction for a halo of 10$^{15}$ $h^{-1}$ M$_{\odot}$, the dashed line
and dotted lines for halos of 10$^{14}$ $h^{-1}$ M$_{\odot}$ and 10$^{13}$ $h^{-1}$ M$_{\odot}$, respectively.  
The substructure function is the number of objects with 
velocities greater than a fraction of the parent halo's 
velocity.
The right panel shows a sample of 5 loose groups
from  Zabludoff  $\&$ Mulchaey (1998a) and the function derived from the LF of
39 Hickson compact groups (Hunsberger, Charlton $\&$ Zaritsky 1998) compared to CDM
predictions (dashed line).}
\end{figure}

\clearpage

\begin{figure}
\plotone{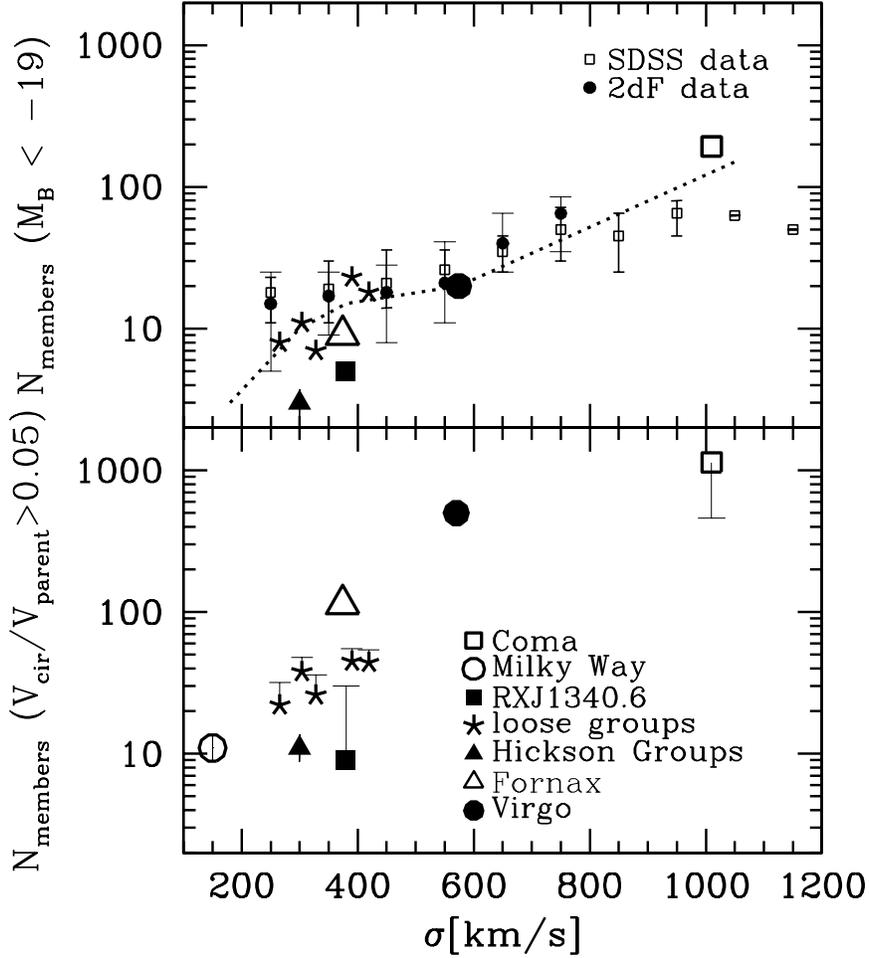}
\caption{The top panel shows the number of members brighter than $M_B < -19$ versus
the velocity dispersion for groups derived from the 2dFGRS (Balogh et al. 2003) and the SDSS
(Desai et al. 2003).  The systems in Fig. 1 are also included.  Note that there are nearly as
many systems in the point labeled ``39 HCGs" as there are groups between 250 and 400
km s$^{-1}$
in the 2dFGRS and SDSS samples.
The dotted line shows what would be expected if the substructure function for
Virgo was universal.   The bottom panel shows the cumulative number of substructures with
circular velocities larger than 5\% of the parent halo's 
circular velocity versus the dispersion of the parent group for the sample shown in
Fig. 1. Data for the Coma cluster of galaxies are from Trentham (1998), but the error bar accounts
results from Mobasher et al. (2003) inferred from a spectroscopic LF.
There is an overall trend with considerable scatter in the region intermediate from galaxies to 
clusters.}

\end{figure}



\end{document}